\documentclass[twocolumn]{aastex63}

\newcommand{\jwst}{{\it JWST}}
\newcommand{\hst}{{\it HST}}
\newcommand{\hz}{high-$z$}
\newcommand{\id}{SC-}
\newcommand{\lya}{{Ly$\alpha$}}
\newcommand{\ha}{H$\alpha$}
\newcommand{\hb}{H$\beta$}
\newcommand{\hi}{H\,{\sc i}}

\newcommand{\oiii}{[O\,{\sc iii]}}
\newcommand{\nii}{[N\,{\sc ii}]}

\newcommand{\buv}{$\beta_{\rm UV}$}
\newcommand{\muv}{$M_{\rm UV}$}
\newcommand{\ewlya}{EW$_0$(Ly$\alpha$)}
\newcommand{\ewha}{EW$_0$(H$\alpha$)}

\newcommand{\fescLya}{$f_{\rm esc}^{\rm Ly\alpha}$}
\newcommand{\fescLyC}{$f_{\rm esc}^{\rm LyC}$}
\newcommand{\ksi}{$\xi_{\rm ion}$}
\newcommand{\ksio}{$\xi_{\rm ion, 0}$}

\newcommand{\Llya}{$L$(\lya)}
\newcommand{\lgLya}{${\rm log}_{10}\,L$(\lya)}

\submitjournal{ApJL}
\shorttitle{\lya\ Galaxies at $z\simeq6$ with \jwst/NIRCam}
\shortauthors{Ning et al.}

\begin{document}

\title{An \ha\ Impression of \lya\ Galaxies at $z\simeq6$ with Deep \jwst/NIRCam Imaging}

\author[0000-0001-9442-1217]{Yuanhang Ning}
\altaffiliation{ningyhphy@mail.tsinghua.edu.cn}
\affiliation{Department of Astronomy, Tsinghua University, Beijing 100084, China}

\author[0000-0001-8467-6478]{Zheng Cai}
\affiliation{Department of Astronomy, Tsinghua University, Beijing 100084, China}

\author[0000-0003-4176-6486]{Linhua Jiang}
\affiliation{Kavli Institute for Astronomy and Astrophysics, Peking University, Beijing 100871, China}
\affiliation{Department of Astronomy, School of Physics, Peking University, Beijing 100871, China}

\author[0000-0001-6052-4234]{Xiaojing Lin}
\affiliation{Department of Astronomy, Tsinghua University, Beijing 100084, China}

\author[0000-0003-0964-7188]{Shuqi Fu}
\affiliation{Kavli Institute for Astronomy and Astrophysics, Peking University, Beijing 100871, China}
\affiliation{Department of Astronomy, School of Physics, Peking University, Beijing 100871, China}

\author[0000-0002-9074-4833]{Daniele Spinoso}
\affiliation{Department of Astronomy, Tsinghua University, Beijing 100084, China}

\begin{abstract}

We present a study of seven spectroscopically confirmed (\lya\ emitting) galaxies at redshift $z\simeq6$ using the \jwst/NIRCam imaging data. These galaxies, with a wide range of \lya\ luminosities, were recently observed in a series of NIRCam broad- and medium-bands. We constrain the rest-frame UV/optical continua and measure the \ha\ line emission of the galaxies using the combination of the \jwst/NIRCam and archival \hst/WFC3 infrared photometry. We further estimate their escape fractions of \lya\ photons (\fescLya) and the production efficiency of ionizing photons (\ksi). Among the sample, 6/7 galaxies have \lya\ escape fractions of ${\lesssim}10\%$, which might be the status for most of star-forming galaxies at $z\simeq6$. One UV-faint \lya\ galaxy with an extremely blue UV slope owns a large value of \fescLya\ reaching ${\simeq}50\%$. These galaxies spread a broad range of \ksi\ over log$_{10}$\,\ksio\,(Hz erg$^{-1}$)~$\sim25.0-26.5$. We find that UV-fainter galaxies with bluer UV continuum slopes likely have higher escape fractions of \lya\ photons. We also find that galaxies with higher \lya\ line emission tend to produce ionizing photons more efficiently. The most \lya-luminous galaxy in the sample has a very high \ksio\ of log$_{10}$\,\ksio\,(Hz erg$^{-1}$)~$>26$. Our results support that \lya\ galaxies may have served as an important contributor to the cosmic reionization. Blue and bright \lya\ galaxies are excellent targets for \jwst\ follow-up spectroscopic observations.

\end{abstract}


\section{Introduction}
The {\it James Webb Space Telescope} (\jwst; \citealt{gardner06}) has begun to explore the very distant Universe, allowing us to gain deep insight on the high-redshift (\hz) objects at the epoch of reionization (EoR). The major sources of reionization are presumably star-forming (SF) galaxies \citep[e.g.,][]{robertson15, finkelstein19, yung20b}. This viewpoint has been conclusively demonstrated by \citet{jiang22}. As a population of SF galaxies, \lya\ emitting galaxies are generally low-mass with low metallicity and dust content \citep[e.g.,][]{haocn18, haro20, santos20}. They should play a non-negligible role to drive the reionization because the processes responsible for the emission of \lya\ and Lyman continuum (LyC) photons relate to each other \citep[e.g.,][]{dijkstra14, verhamme15, deBarros16, dijkstra16} while \lya\ escape fraction (\fescLya) generally exceeds LyC escape fraction \citep[\fescLyC; e.g.,][]{izotov16, izotov20, leitherer16, shapley16, verhamme17, flury22}.

Large ground-based telescopes and {\it Hubble Space Telescope} (\hst) helped us to find a large number of \lya-emitting galaxies with redshift reaching $z\gtrsim6-7$, corresponding to the end of EoR. Most of them are \lya\ emitters (LAEs) selected by the narrowband (\lya) technique \citep[e.g.,][]{kashikawa06, kashikawa11, hu10, shibuya18b, taylor21}. The rest of them are Lyman-break galaxies (LBGs) selected by the dropout technique and identified by \lya\ lines \citep[e.g.,][]{steidel96b, jones12, inami17, pentericci18b}. The LAEs and LBGs (with \lya\ lines) are probably indistinguishable in terms of their intrinsic properties such as age, stellar mass, and star formation rate \citep[SFR; e.g.,][]{dayal12, jiang16a, delavi20}. We thus call both of them as \lya\ galaxies in the following text.

To understand how SF galaxies contribute to the ionizing photon budget, their rest-frame optical properties (continua and line emission) are necessary to be constrained. However, this task is difficult to execute before \jwst\ era, especially for $z\simeq6$ galaxies. For example, due to a lack of near-/mid- infrared (IR) bands, it is challenging to break the degeneracy between prominent nebular emission from young galaxies and strong Balmer breaks from old galaxies \citep[e.g.][]{schaerer09, jiang16a}. On the other hand, even if the galaxies are spectroscopically confirmed (by \lya\ line for instance) at $z\simeq6$, the optical emission lines (mainly \oiii+\hb\ and \ha+\nii) simultaneously boost the IRAC1 and IRAC2 channels of {\it Spitzer Space Telescope} \citep[e.g.][]{faisst16, harikane18b, stefanon21}, leaving lines and continua coupled together. Such problems are being well solved in the current \jwst\ era.

For building a large and homogeneous sample of \hz\ galaxies, we have carried out the Magellan M2FS spectroscopic survey to identify LAEs and LBGs at $z\simeq6$ (\citealt{jiang17, ning20, ning22}; Fu et al. in preparation). A fraction of them will be covered by the upcoming \jwst\ imaging survey, such as COSMOS-Webb \citep[GO 1727;][]{cosmosWebb, casey22} and PRIMER \citep[GO 1837;][]{primer}. The multiple IR bands can reveal their individual properties in detail. Recently, one of our identified LBGs has been covered by parallel \jwst/NIRCam imaging of PRIMER MIRI observations. In this work, we compare it with spectroscopically confirmed galaxies at $z\simeq6$ from from previous literatures \citep{pentericci18b} to give a pilot investigation on the \ha\ properties of luminous \lya\ galaxies.

This paper is organized as follows. In Section 2, we briefly present the sample of \lya\ galaxies, \jwst/NIRCam imaging observations, data reduction, and photometry. In Section 3, we give the measurement results of the \lya, ultraviolet (UV), and \ha-related properties of the galaxy sample. We discuss their \lya\ escape fractions and ionizing photon production rates in Section 4. We summarize this work in Section 5. Throughout the paper, we use a standard flat cosmology with $H_0=\rm{70\ km\ s^{-1}\ Mpc^{-1}}$, $\Omega_m=0.3$ and $\Omega_{\Lambda}=0.7$. All magnitudes refer to the AB system \citep{oke74}.

\section{Sample and Data}
In this section, we describe our sample of \lya\ galaxies at $z\simeq6$, \jwst/NIRCam imaging observations, data reduction, and photometry. We summarize the sample information in Table~\ref{sample}.

\begin{figure}
\epsscale{1.15}
\centering
\plotone{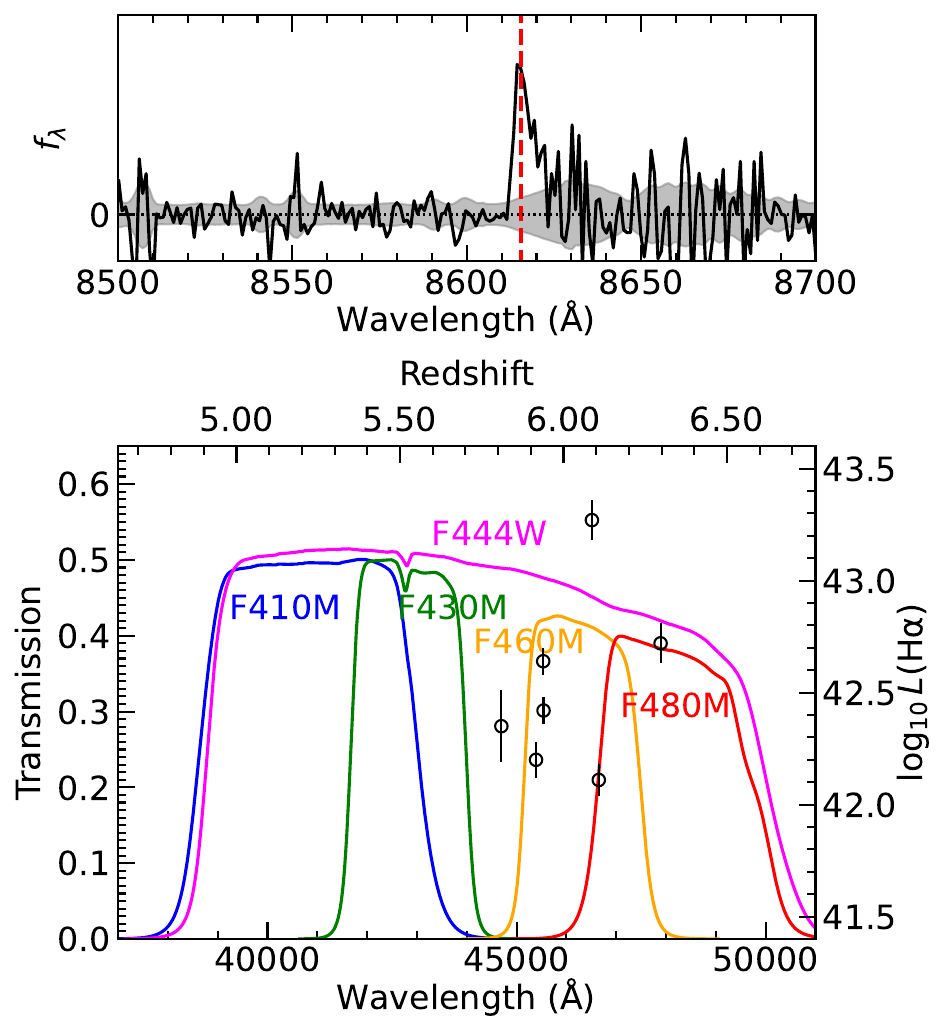}
\caption{{\it Upper panel:} The spectrum of the \lya\ galaxy (\id1 with $z_{\rm Ly\alpha}=6.087$) confirmed by our Magellan M2FS spectroscopic survey. The vertical dashed line marks the observed \lya\ wavelength. The shaded region represents $\pm1\sigma$ noise level. 
{\it Lower panel:} The redshift-\ha\ luminosity distribution of the sample in this work and the transmission curves of the five \jwst/NIRCam filters. The highest datapoint corresponds to our confirmed galaxy (\id1) shown in the upper panel.
\label{zdist}}
\end{figure}

\renewcommand{\arraystretch}{1}
\begin{table*}[t]
\caption{Basic information and photometry of the \lya\ galaxy sample at $z\simeq6$}
\centering
\begin{tabular}{c|ccccccc}
\hline\hline
                          ID &              SC-1 &              SC-2 &              SC-3 &              SC-4 &              SC-5 &              SC-6 &              SC-7 \\
\hline                RA (J2000.0) &       02:17:43.25 &       02:17:48.31 &       02:17:25.12 &       03:32:28.19 &       03:32:36.47 &       03:32:38.28 &       03:32:39.06 \\
               Dec (J2000.0) &     $-$05:06:47.5 &     $-$05:10:31.7 &     $-$05:11:35.0 &     $-$27:48:18.7 &     $-$27:46:41.4 &     $-$27:46:17.2 &     $-$27:45:38.7 \\
 Redshift $z_{\rm Ly\alpha}$ &             6.087 &             5.810 &             6.297 &             5.939 &             5.938 &             6.108 &             5.916 \\
               CANDELS field &               (UDS)\tablenotemark{a} &               UDS &               UDS &           GOODS-S &           GOODS-S &           GOODS-S &           GOODS-S \\
                  JWST ObsID &       PRIMER-o022 &       PRIMER-o022 &       PRIMER-o014 &            UDF-MB &            UDF-MB &            UDF-MB &            UDF-MB \\
                 F090W (mag) &  26.58 $\pm$ 0.07 &  26.02 $\pm$ 0.04 &               ... &               ... &               ... &               ... &               ... \\
                 F105W (mag) &               ... &               ... &               ... &  26.15 $\pm$ 0.06 &  25.76 $\pm$ 0.03 &  26.31 $\pm$ 0.03 &  27.64 $\pm$ 0.07 \\
                 F115W (mag) &  26.56 $\pm$ 0.07 &  25.56 $\pm$ 0.03 &               ... &               ... &               ... &               ... &               ... \\
                 F125W (mag) &               ... &  25.56 $\pm$ 0.09 &  26.39 $\pm$ 0.17 &  26.15 $\pm$ 0.07 &  25.70 $\pm$ 0.02 &  26.35 $\pm$ 0.02 &  27.73 $\pm$ 0.06 \\
                 F150W (mag) &  26.76 $\pm$ 0.08 &  25.30 $\pm$ 0.02 &               ... &               ... &               ... &               ... &               ... \\
                 F160W (mag) &               ... &  25.49 $\pm$ 0.10 &  26.51 $\pm$ 0.16 &  26.04 $\pm$ 0.07 &  25.70 $\pm$ 0.02 &  26.32 $\pm$ 0.02 &  28.06 $\pm$ 0.08 \\
                 F182M (mag) &               ... &               ... &               ... &  26.32 $\pm$ 0.01 &  26.20 $\pm$ 0.01 &  26.32 $\pm$ 0.01 &  27.46 $\pm$ 0.02 \\
                 F210M (mag) &               ... &               ... &               ... &  26.21 $\pm$ 0.01 &  26.07 $\pm$ 0.01 &  26.28 $\pm$ 0.01 &  27.15 $\pm$ 0.02 \\
                 F200W (mag) &  26.57 $\pm$ 0.07 &  25.22 $\pm$ 0.02 &               ... &               ... &               ... &               ... &               ... \\
                 F277W (mag) &  26.20 $\pm$ 0.04 &  24.80 $\pm$ 0.01 &  25.97 $\pm$ 0.03 &               ... &               ... &               ... &               ... \\
                 F356W (mag) &  25.06 $\pm$ 0.01 &  24.57 $\pm$ 0.01 &  25.21 $\pm$ 0.01 &               ... &               ... &               ... &               ... \\
                 F410M (mag) &  25.93 $\pm$ 0.07 &  24.66 $\pm$ 0.02 &  26.26 $\pm$ 0.09 &               ... &               ... &               ... &               ... \\
                 F444W (mag) &  24.73 $\pm$ 0.02 &  24.53 $\pm$ 0.02 &  25.67 $\pm$ 0.04 &               ... &               ... &               ... &               ... \\
                 F430M (mag) &               ... &               ... &               ... &  26.42 $\pm$ 0.09 &  26.03 $\pm$ 0.06 &  26.27 $\pm$ 0.08 &  27.67 $\pm$ 0.34 \\
                 F460M (mag) &               ... &               ... &               ... &  25.06 $\pm$ 0.03 &  24.56 $\pm$ 0.02 &  25.51 $\pm$ 0.04 &  25.80 $\pm$ 0.07 \\
                 F480M (mag) &               ... &               ... &               ... &  26.31 $\pm$ 0.09 &  25.92 $\pm$ 0.06 &  25.79 $\pm$ 0.05 &            $>$27.62\tablenotemark{b}
\\\hline
\end{tabular}
\label{sample}
\tablenotemark{a}{The parenthesis indicates that \id1 is located close to the CANDELS-UDS imaging region.}\\
\tablenotemark{b}{This value correspond to a $3\sigma$ upper limit.}
\end{table*}

\begin{figure*}[t]
\centering
\gridline{\fig{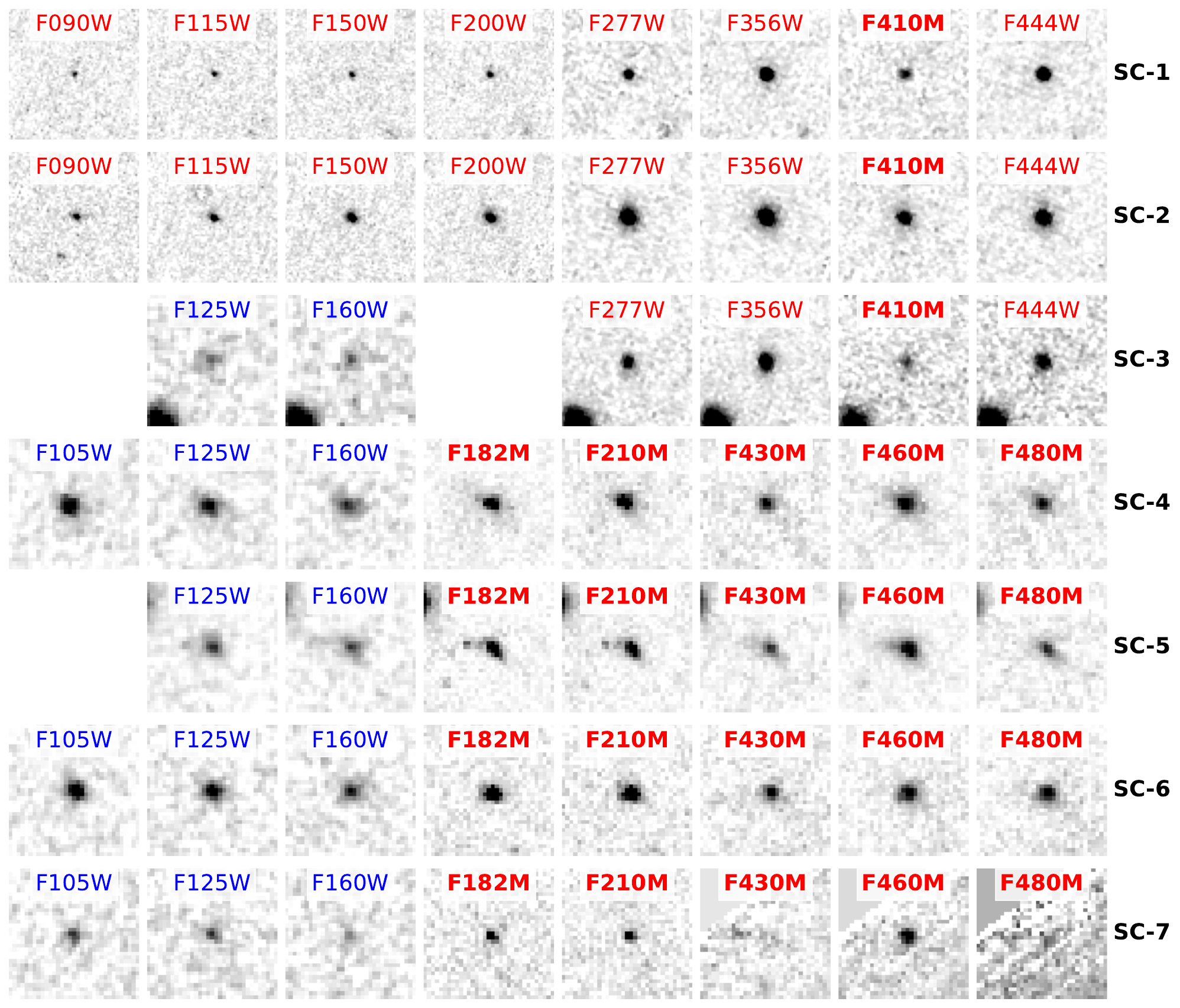}{.62\textwidth}{(a)}
\fig{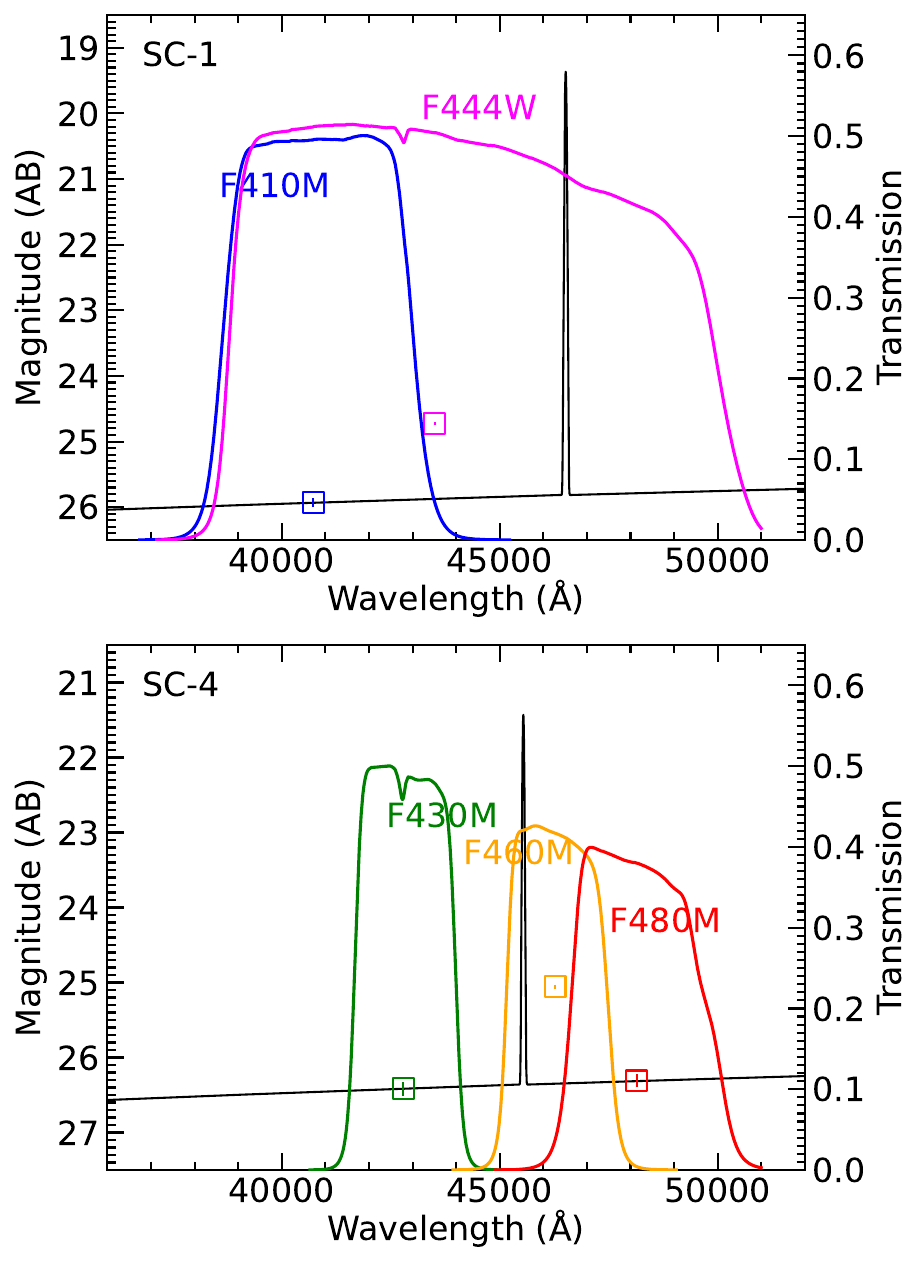}{.38\textwidth}{(b)}}
\caption{\textbf{(a)} Thumbnail images of the \lya\ galaxies at $z\simeq6$ in this work. Their ID names are marked at the right end of each row. The size of the images is 2\arcsec$\times$2\arcsec\ (north is up and east to the left). The corresponding band is marked at the top of each thumbnail image (\hst/WFC3 bands in blue and \jwst/NIRCam bands in red). The medium bands are shown in the bold face.
\textbf{(b)} Illustration of measuring \ha+\nii\ line flux of the seven sources (shown in the left figure) including three sources covered by the PRIMER survey (the upper panel gives an example) and four sources covered by the UDF-MB survey (the lower panel gives an example). The boxes with errorbars are photometry in the NIRCam LW bands. They have the same colors as the transmission curves of the five NIRCam filters. The black lines represent the power-law optical continua and gaussian profiles (FWHM = 300 km s$^{-1}$) of \ha\ lines which boost the corresponding bands.
\label{cuts}
}
\end{figure*}

\subsection{Sample of \lya\ Galaxies at $z\simeq6$}

The sample includes seven spectroscopically confirmed galaxies at $z\simeq6$. The first one (\id1) is confirmed at redshift $z=6.087$ with a strong \lya\ line by our spectroscopic survey (see the upper panel of Figure~\ref{zdist}; Fu et al. in preparation). In this survey, we carried out spectroscopic observations using the fiber-fed, multi-object spectrograph Michigan/Magellan Fiber System (M2FS; \citealt{mateo12}) on the 6.5 m Magellan Clay telescope. The science goal is to build a large and homogeneous sample of \hz\ galaxies (see \citealt{jiang17} for an overview of the program), including LAEs at $z\approx5.7$ and 6.6 \citep{ning20, ning22}, and LBGs at $5.5<z<6.8$ (Fu et al. in preparation). These \hz\ galaxies are located in the famous fields including the Subaru {\it XMM-Newton} Deep Survey (SXDS), the Extended {\it Chandra} Deep Field-South (ECDFS), A370, COSMOS, and SSA22. The total sky area is around $2$ deg$^2$. \id1 has an estimated \lya\ equivalent width of \ewlya\ $\gtrsim100$ \AA. It is thus one of the largest-\ewlya\ LBGs in a wide area covered by our spectroscopic survey. 

The rest six sources (\id2--7) are from a previous work, \citet[][hereafter P18]{pentericci18b}. They used VLT/FORS2 to conduct the CANDELSz7 survey, an ESO Large Program, to spectroscopically confirm SF galaxies at $z\gtrsim6$ in three \hst\ CANDELS Legacy fields (GOODS-South, UDS, and COSMOS). \id2 and \id3 locate in the CANDELS UDS field while \id4--7 locate in the CANDELS GOODS-South (GOODS-S) field. Among them, \id2 was detected with no emission line and a continuum discontinuity which is interpreted as a \lya\ break (see P18). Due to the limited sensitivity of instruments and depth of observations, we treat this source as a \lya\ galaxy with an upper limit of \ewlya\ from P18. Figure~\ref{zdist} shows redshift distribution of the sample in the lower panel.

\subsection{Imaging Data}

The three sources \id1--3 in/around the CANDELS UDS field are covered by the \jwst\ Cycle-1 program (GO 1837), Public Release IMaging for Extragalactic Research (PRIMER; \citealt{primer}). The PRIMER survey pays attention to the two key equatorial \hst\ CANDELS Fields (COSMOS and UDS) by delivering 10-band NIRCam+MIRI imaging observations. It owns the parallel NIRCam imaging in eight bands including F090W, F115W, F150W, and F200W in the short wavelength (SW), and F277W, F356W, F444W, and F410M in the long wavelength (LW). Note that the F410M$-$F444W color is a good indication of \ha\ emission of the SF galaxies at $z\sim5-6.6$. \id1--3 have been imaged by PRIMER \#14 and \#22 observations with individual exposure lengths of ${\sim}1.9$ hours in the SW bands and ${\sim}0.5$ hours in the LW bands.

The other four sources \id4--7 are covered by the \jwst\ Cycle-1 program (GO 1963), UDF medium-band survey (UDF-MB; \citealt{udfMB}). The UDF-MB survey images the (Hubble) Ultra Deep Field (UDF) with a single NIRCam pointing (field of view $\sim2\times2\arcmin.2\times2\arcmin.2$) in a series of medium-bands including F182M, F210M, F430M, F460M, and F480M (NIRISS F430M and F480M in parallel). Total integration time reaches ${\sim}7.8$ hours for each F182M, F210M, and F480M and ${\sim}3.9$ hours for each F430M and F460M. In Figure~\ref{zdist}, we plot the transmission curves of the five \jwst/NIRCam LW filters to compare and illustrate the \ha\ locations of the galaxies in the observed-wavelength frame.

Except \id1 and 2, other sources are not covered by enough \jwst\ SW bands especially \id3 is located in the gap region of NIRCam SW imaging. We thus utilize the archival \hst/WFC3 (Wide Field Camera 3) near-IR imaging data from the CANDELS program \citep{candels1, candels2, candelsGDScat, candelsUDScat}. We download the data products provided by the High Level Science Products\footnote{\url{https://archive.stsci.edu/hlsp/candels}}. \id2 and \id3 have \hst/WFC3 F125W and F160W observations while \id1 is a little bit outside the CANDELS UDS region. \id4--7 in the CANDELS GOODS-S region have \hst/WFC3 observations in three near-IR bands (F105W, F125W and F160W).

\subsection{Data Reduction and Photometry}

We reduced the NIRCam imaging data with the standard \jwst\ pipeline\footnote{\url{https://github.com/spacetelescope/jwst}} (v1.7.2) up to stage 2 using the reference files “jwst\_0999.pmap” for PRIMER and “jwst\_1008.pmap” for UDF-MB. Then we use the \texttt{Grizli}\footnote{\url{https://github.com/gbrammer/grizli}} reduction pipeline to process the output images. \texttt{Grizli} mitigates 1/f noises and mask the “snowball” artifacts from cosmic rays \citep{rigby22}.  It further converts the world coordinate system (WCS) information in the headers to the SIP format for each exposure so that images can be drizzled and combined with Astrodrizzle\footnote{\url{https://drizzlepac.readthedocs.io/en/latest/astrodrizzle.html}}. For the SW and LW images, the WCS of final mosaics are registered based on the catalogs of DESI Legacy Imaging Surveys Data Release 9 and the pixel scale was resampled to 0.03\arcsec\ with pixfrac $=0.8$. We also subtract an additional background on the final mosaics. Figure~\ref{cuts}a shows the thumbnail images of the sample in a series of the \jwst/NIRCam (and/or \hst/WFC3 near-IR) bands.

We run \texttt{SExtractor} \citep{ber96} to perform photometry in the \jwst/NIRCam multi-band images. The aperture has a radius of triple FWHMs of the point-spread function (PSF) in each wavelength band. The aperture correction is calculated from the PSF in each band. We first obtain initial measurements by matching the output catalogs to the targets within a distance tolerance of a FWHM. For each target, we select its brightest band to feed the detection image. Specifically, we adopt the F444W band for three sources covered by the PRIMER survey and the F460W band for the four sources covered by the UDF-MB survey because their \ha\ lines boost these bands. We then rerun \texttt{SExtractor} in the dual image mode with the detection images. For each measurement image, we also adopt an aperture with a radius of triple PSF FWHMs in this band. Only for the \id7 source, we use a radius of 1.5 PSF FWHM, in order to minimize the amount of abnormal pixels within the photometric aperture, caused by the fact that SC-7 is located very close to the image edge. Table \ref{sample} lists the multi-band photometry results of the galaxy sample.

\floattable
\begin{deluxetable}{ccccccccccc}[t]
\tablecaption{Measured properties of the \lya\ galaxy sample at $z\simeq6$.
\label{measure}}
\tablewidth{0pt}
\centering
\tablehead{
   \colhead{ID} & \colhead{$z_{\rm Ly\alpha}$} &\colhead{$\beta_{\rm UV}$} & \colhead{$M_{\rm UV}$} & \colhead{${\rm log}_{10} L$(\lya)} & \colhead{\ewlya} & \colhead{$f_{\rm esc}$({\lya})} & \colhead{${\rm log}_{10} L$(\ha)} & \colhead{\ewha} & \colhead{${\rm log}_{10}\xi_{\rm ion, 0}$} & \colhead{SFR(\ha)} \\
   \colhead{} & \colhead{} &\colhead{} & \colhead{} & \colhead{($\rm erg\ s^{-1}$)} & \colhead{(\AA)} & \colhead{} & \colhead{($\rm erg\ s^{-1}$)} & \colhead{(\AA)} & \colhead{(Hz erg$^{-1}$)} & \colhead{(M$_{\odot}$ yr$^{-1}$)} \\
   \colhead{(1)} & \colhead{(2)} &\colhead{(3)} & \colhead{(4)} & \colhead{(5)} & \colhead{(6)} & \colhead{(7)} & \colhead{(8)} & \colhead{(9)} & \colhead{(10)} & \colhead{(11)}
   }
\startdata
 SC-1 &  6.087 &   -2.03 &  -20.11 &   43.16 &    147 &  0.066 $\pm$ 0.014 &  43.27 $\pm$ 0.09 &  2756 $\pm$ 596 &  26.45 $\pm$ 0.19 &  100 $\pm$ 22 \\
 SC-2 &  5.810 &   -1.46 &  -21.06 &  $<$41.92 &     $<$4 &              $<$0.03 &  42.35 $\pm$ 0.16 &   113 $\pm$ 40  &  25.15 $\pm$ 0.31 &    12 $\pm$ 4 \\
 SC-3 &  6.297 &   -2.53 &  -20.46 &   42.54 &     23 &  0.055 $\pm$ 0.011 &  42.72 $\pm$ 0.09 &   994 $\pm$ 204 &  25.77 $\pm$ 0.18 &    28 $\pm$ 6 \\
 SC-4 &  5.939 &   -1.73 &  -20.51 &   42.44 &     20 &  0.088 $\pm$ 0.013 &  42.42 $\pm$ 0.06 &   663 $\pm$ 94  &  25.45 $\pm$ 0.12 &    14 $\pm$ 2 \\
 SC-5 &  5.938 &   -1.88 &  -20.93 &   42.15 &      7 &  0.027 $\pm$ 0.004 &  42.64 $\pm$ 0.06 &   767 $\pm$ 102 &  25.49 $\pm$ 0.12 &    23 $\pm$ 3 \\
 SC-6 &  6.108 &   -2.00 &  -20.39 &   42.10 &     10 &  0.082 $\pm$ 0.013 &  42.11 $\pm$ 0.07 &   267 $\pm$ 42  &  25.18 $\pm$ 0.13 &     7 $\pm$ 1 \\
 SC-7 &  5.916 &   -3.01 &  -19.08 &   42.94 &    186 &  0.470 $\pm$ 0.085 &  42.20 $\pm$ 0.08 &  1261 $\pm$ 227 &  25.79 $\pm$ 0.16 &     8 $\pm$ 2 
\enddata
\centering
\end{deluxetable}

\section{Results}
In this section, we give the measured results of UV and \ha\ quantities of the galaxy sample. Their UV properties are derived from \jwst/NIRCam SW bands and/or \hst/WFC3 near-IR bands. We utilize the NIRCam medium-bands to constrain their (rest-frame) optical continuum. We then measure their \ha\ flux by combining the corresponding LW broad- or medium-bands. We further obtain their \ha-related properties including the \lya\ escape fraction and the ionizing photon production efficiency. The results are listed in Table~\ref{measure}.

\subsection{UV Continua}

We measure the UV continuum of the galaxies with the commonly used method \citep[e.g.,][]{pentericci18b, jiang20a}. As in these works, we assume a power-law form for the UV continuum of each source, i.e.: $f_\lambda \propto \lambda^{\beta}$. As we work in AB magnitude units, we fit a linear relation $m_{\rm AB} \propto (\beta+2)\times{\rm log}(\lambda)$ to the SW photometric data, from which we obtain the UV continuum slope \buv\ and the absolute UV magnitude \muv\ at the rest-frame wavelength 1500\AA. 

For a galaxy at $z\simeq6$, its observed \lya\ line locates in the wavelength range of the Subaru/$z'$ and \jwst/F090W bands. The corresponding broad-band flux usually differs from the flux level of UV continuum due to the \lya\ emission or break (IGM absorption bluewards of \lya). So in the measurements for UV continua, we abandon the F090W photometric data for \id1 and \id2. We then subtract the fit power-law UV continuum from the $z'$- or F090W-band photometry to constrain \lya\ flux and \ewlya\ for \id1. The IGM continuum absorption blueward of \lya\ line is considered in the computation \citep{madau95}. For \id3--7, the CANDELz7 galaxies, we directly adopt the observed \lya\ flux given by P18 to compute the \lya\ luminosity and \ewlya\ with our obtained power-law UV continua. For \id2 which is undetected in \lya, we use the upper limit of \ewlya\ given by P18 to compute its \lya\ flux. Note that \lya\ flux may be slightly underestimated due to the potential \lya\ emission from the circumgalactic medium \citep[e.g.,][]{cai19, wu20}.

\subsection{\ha\ Line Emission}

We combine the medium- and broad-bands (covering the rest-frame optical wavelength at $z\simeq6$) to estimate the flux and EW of \ha\ emission lines. In Figure~\ref{cuts}, the (red) color of F410M$-$F444W and F430M$-$F460M clearly show the flux excess due to strong \ha\ lines. In the F444W broad-band which covers the F410M, F430M, F460M, and F480M medium-bands, \ha+\nii\ lines dominate the flux estimation \citep[e.g.,][]{af03}. We thus ignore other optical lines except \ha+\nii. We assume that \nii\ contribute line flux at the \ha\ wavelength due to the small wavelength difference relative to the wavelength range of ${>}4\ \micron$. We also assume that the emission line has a gaussian profile with FWHM = 300 km s$^{-1}$ (see the black lines in Figure \ref{cuts}b).

\id4--7 are covered by the UDF-MB survey with five NIRCam medium-bands. For \id4, 5, and 7, we use the F430M and F480M magnitude ($3\sigma$ upper limit for \id7) to constrain the rest-frame optical continuum with a power-law form $f_{\nu}\propto\nu^{\alpha}$ because the \ha+\nii\ emission only fall into the F460M band. For \id6, the \ha+\nii\ emission fall into the F460M and F480M bands. We thus match the continuum plus line model to the three LW medium-band photometric data. Figure~\ref{cuts}b illustrates the above procedure in the lower panel. For these four sources, we obtain similar power-law indices with a median value of $\alpha\sim-0.6$. \id1--3 are covered by the PRIMER NIRCam multi-band (7 broad + 1 medium) observations. As no strong nebular lines fall into the F410M band, we use the F410M magnitude to constrain the (rest-frame) optical continuum and the F410M$-$F444W color to estimate \ha+\nii\ flux. We start with a power-law continuum with an index of $\alpha_0=-0.6$ (from the other four sources) to match the F410M flux density. Then we integrate the known continuum plus unknown line emission weighted by the F444W filter transmission curve to match the F444M flux density and compute the \ha+\nii\ flux. We also vary the continuum slope $\alpha$ in a reasonable range of $\alpha_0\pm0.5$ to obtain the deviations of the measured line flux which would be included into the errors of the final values. Figure~\ref{cuts}b illustrates the above procedure in the upper panel.

After estimating the line flux, we assume that \ha\ accounts for 85\% of the \ha+\nii\ flux, which is similar to previous studies \citep[e.g.,][]{rasappu16, faisst19, sun22b}. To be conservative, we also feed the 10\% flux into the error of the final \ha\ flux. With the measured flux of \ha\ line, we obtain the SFR(\ha) using the canonical \ha-SFR calibration relation \citep[listed in Column 11 of Table~\ref{measure};][]{hao11, murphy11, ke12}. We further compute the \ewha\ with the rest-frame optical continuum level. The results are plotted in the second and third rows of Figure~\ref{propt}. Note that the \ha\ flux and EW may be underestimated because the optical continuum is overestimated due to the existence of some faint optical lines. But such an underestimation is supposed to be included into the enlarged measurement errors. \ewha\ indicates the specific SFR (sSFR) of galaxies. Our $z\simeq6$ sample spread a larger range of \ewha\ than the low-$z$ LAEs (\citealt{matthee21}) and local analogs (\citealt{yang17b}). Recently, \citet{sun22b} serendipitously found a sample of strong \ha/\oiii\ emitters in the \jwst/NIRCam wide-field slitless spectroscopy (WFSS) data. Our median \ewha\ is twice higher than theirs because in our sample the 6/7 galaxies emitting \lya\ lines are supposed to have higher sSFR while the SF galaxies of \citet{sun22b} are found based on \ha/\oiii\ detections. \id7 is undetected in \lya\ and its \ewha\ is similar to the lowest one of the sample in \citet{sun22b}.

\subsection{\ha-related Properties}

\subsubsection{Escape Fraction of \lya\ Photons}
We estimate the escape fraction of \lya\ photons (\fescLya) for the sample. With the obtained \ha\ flux, we adopt the canonical ratio $L({\rm Ly\alpha})/L({\rm H\alpha})=8.7$ \citep[e.g.,][]{henry15} to calculate the intrinsic \lya\ flux and obtain: 
\begin{eqnarray}
f_{\rm esc}^{\rm Ly\alpha} = \frac{L_{\rm obs}({\rm Ly\alpha})}{L({\rm H\alpha})\times8.7}
\end{eqnarray} under the assumption of case-B recombination in $T_e=10^4 K$ \citep{agn2}. We also apply a dust correction using the reddening law of \citet{calzetti00}. We can not well constrain the extinction $E(B-V)$ for nebulae due to a lack of the Balmer decrement (\ha/\hb) information. Thus, we perform SED fitting using \texttt{BAGPIPES} \citep{carnall18} and obtain $A_V=0.8$ for \id2 thanks to the abundance of multi-band photometric data for this source. For others, we adopt a lower and modest value of $A_V=0.4$ as a reasonable assumption because they have higher \ewlya\ with lower dust content. Note that $A_V$ is supposed to be smaller for the six galaxies but the difference is only $\lesssim0.1$ dex for the computed \fescLya. The \fescLya\ results are listed in the Column 7 of Table~\ref{measure}. We can see that 6/7 galaxies have \fescLya~$\lesssim10\%$ even though they spread over a large range of \ewlya\ and \ewha. This may imply the upper limit of \fescLya\ for most of galaxies at $z\simeq6$. Note that \fescLyC\ is supposed to be smaller than the \fescLya\ \citep[e.g.,][]{dijkstra16, izotov20}. \id7 has the largest value of \fescLya\ reaching ${\simeq}50\%$ while it is relatively faint in the rest-frame UV (\muv~$\simeq-19$) with an extremely blue UV slope (\buv~$\simeq-3$). In Section 4.1, we discuss the relevant, potential trends.

\subsubsection{Production Efficiency of Ionizing Photons}
With the measured flux of UV continua and \ha\ line, we estimate the Hydrogen ionizing photon production efficiency \ksi\ by
\begin{eqnarray}
   \xi_{\rm ion} = \frac{\dot N_{\rm ion}}{L_{\nu}^{\rm UV}},
\end{eqnarray} where $\dot N_{\rm ion}$ (s$^{-1}$) is the intrinsic production rate of Hydrogen ionizing photons from stellar populations and $L_{\nu}^{\rm UV}$ (erg~s$^{-1}$~Hz$^{-1}$) is the (mono-chromatic) UV continuum luminosity per photon frequency which can be derived from the above measured \muv. \ksi\ intrinsically depends on the assumed stellar-population model \citep[e.g.,][]{robertson13, eldridge17, yung20a}. $\dot N_{\rm ion}$ can be computed from \ha\ emission by 
\begin{eqnarray}
\dot N_{\rm ion} = \frac{L({\rm H\alpha})}{1-f_{\rm esc}^{\rm LyC}} \times 7.35\times10^{11}\ {\rm erg^{-1}}, 
\end{eqnarray} in the ($T_e=10^4 K$) case-B recombination \citep{kennicutt94, lh95, madau98}. We then obtain the production efficiency of ionizing photons which do not escape from the galaxy, \ksio\ assuming \fescLyC\ = 0.

\begin{figure*}[t]
\centering
\includegraphics[angle=0, width=1\textwidth]{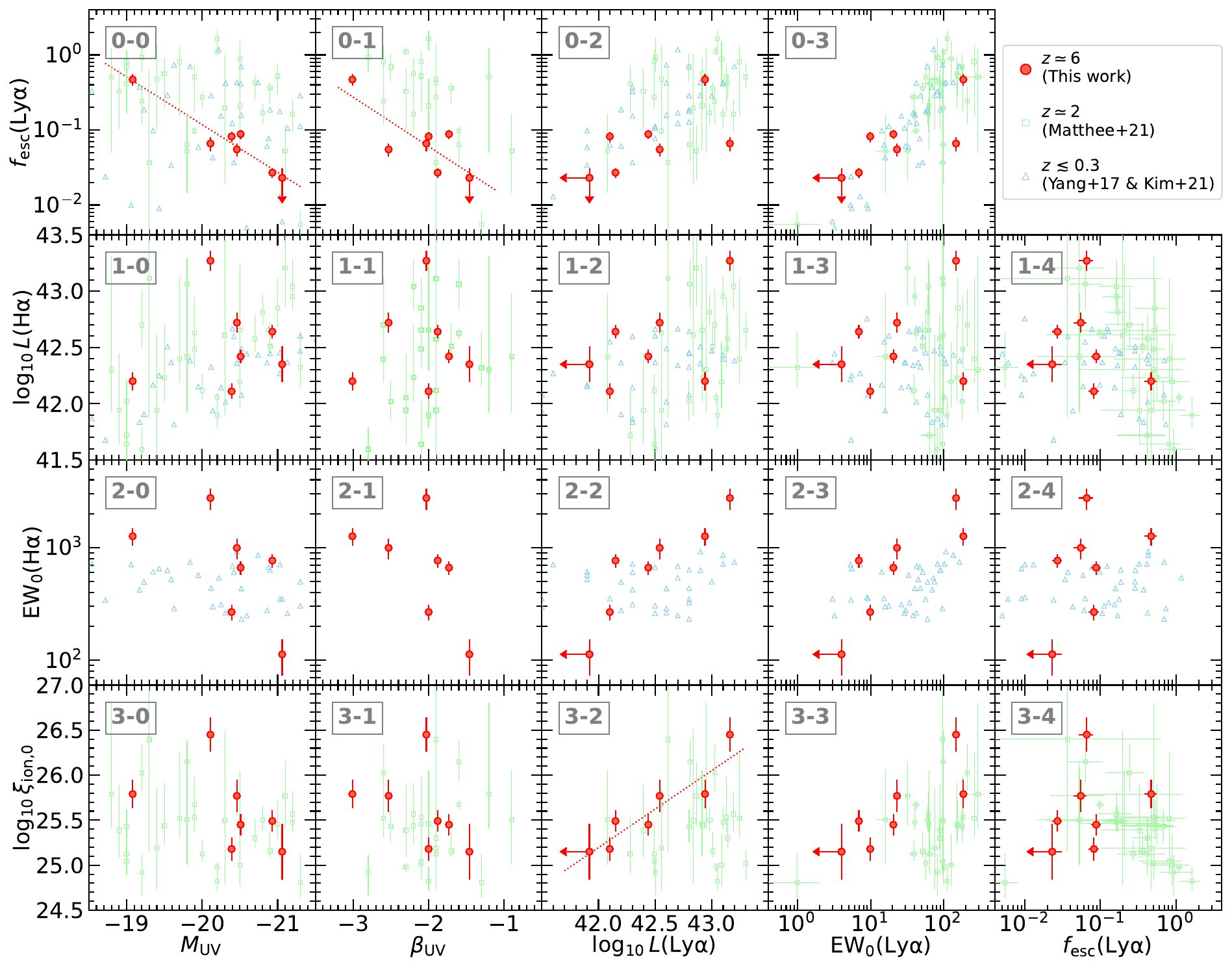}
\caption{Comparison of the measured \ha-related properties with the \lya\ and UV properties. From top to bottom, we show the \lya\ escape fraction, \ha\ luminosity, rest-frame \ha\ EW, and ionizing photon production efficiency. From left to right, we show the rest-frame UV magnitude, UV slope, \lya\ luminosity, and rest-frame \lya\ EW. We also compare the \lya\ escape fraction with the measured \ha\ properties in the rightmost column. The red circles indicate the $z\simeq6$ galaxies measured by this work. The green squares represent the $z\sim2$ LAEs from \citet{matthee21}. The blue triangles correspond to the local Green Pea galaxies from \citet{yang17b} and \citet{kimk21}. The red dotted lines indicate the best linear fits to the measured quantities (red circles).
\label{propt}}
\end{figure*}

The (dust-uncorrected) results are shown in the downmost row of Figure~\ref{propt}. Note that the dust-corrected values are lower by ${\lesssim}0.1$ dex assuming the canonical stellar/nebular extinction ratio of 0.44 which is obtained from local starbursts \citep{calzetti00}. The \ksio\ results are listed in the Column 10 of Table~\ref{measure}. Our obtained \ksio\ distribute over a broad range of log$_{10}$\,\ksio~$\sim25.0-26.5$. The median value is consistent with that of a large UV-faint galaxy sample from \citet{2022arXiv221112548P}. This median is also close to that of a \ha-emitter sample from \citet{sun22b}. Among our sample, \id1 is the most luminous in \lya\ and has the highest \ksio\ reaching log$_{10}$\,\ksio~$\sim26.5$ while \id2 is not detected in \lya\ and has the lowest \ksio. We further give an extensive discussion in next section.

\section{Discussion}
Thanks to the excellent capability of \jwst\ \citep{rigby22}, we have measured the \ha-related properties of the SF galaxies at $z\simeq6$. In this section, we discuss the obtained properties of the galaxy sample. We compare our results with previous studies in Figure~\ref{propt}. The \ha-related properties are assigned by rows and \lya/UV quantities are assigned by columns. The \fescLya\ is assigned again at the rightmost column to compare with the \ha-related properties.

\subsection{\fescLya-UV Correlation}
We compare the obtained \fescLya\ in the upmost row and rightmost column of Figure~\ref{propt}. The measured results show that \fescLya\ positively correlates with \Llya\ and \ewlya, and negatively correlates with \muv\ and \buv\ (in the upmost row). The positive trends are natural because \fescLya\ is inferred through \lya\ flux, which is similar to those found from the low-$z$ samples \citep[e.g.,][]{hayes14, yang17b, matthee21}. \citet{yang17b} (and \citealt{kimk21}) use a statistical sample of local Green Pea galaxies as \hz\ analogs to reveal that. \citet{matthee21} use a LAE sample at $z\sim2$ and also found such a relation. For the negative trends of \fescLya\ changing as UV properties, we obtain a linear relation between \fescLya\ and \muv,
\begin{equation}
{\rm log_{10}}\,f_{\rm esc}^{\rm Ly\alpha} = (0.64 \pm 0.08)\,M_{\rm UV} + (11.8 \pm 1.7),
\label{eq1}
\end{equation} shown as a red dotted line in panel 0-0; we also obtain another linear relation between \fescLya\ and \buv,
\begin{equation}
{\rm log_{10}}\,f_{\rm esc}^{\rm Ly\alpha} = (-0.66 \pm 0.23)\,\beta_{\rm UV} - (2.5 \pm 0.5),
\label{eq2}
\end{equation} shown as a red dotted line in panel 0-1.
The \fescLya-\muv\ relation seems to exist in the current \hz\ sample, although the sign is weak for the two samples of \lya\ galaxies at lower redshifts. \citet{chisholm22} found that \fescLyC\ increases for fainter \muv\ using a sample of LyC-leaking SF galaxies at $z\simeq0.3$, which is overall lower than our obtained \fescLya-\muv\ trend. Such difference is consistent with the scenario of \fescLyC\ $\lesssim$ \fescLya\ \citep[e.g.,][]{deBarros16, dijkstra16}.

The \fescLya-\buv\ trend may exist at $z\simeq6$ like at $z\sim2$ \citep[e.g.,][]{2022arXiv221100041S, 2022arXiv221112548P}. The reason can be simply interpreted as that bluer galaxies have younger stellar population and/or lower dust content. They emit harder ionizing photons suffering lower extinction, which accounts for higher escape fraction of \lya\ photons. The $z\simeq6$ trend is steeper, which is consistent with the current consensus that galaxies at higher redshift are bluer \citep[e.g.,][]{bouwens14, jiang20a}. \citet{chisholm22} shows that \fescLyC\ scales strongly with \buv\ for the LyC-leaker sample at $z\simeq0.3$, while those with \buv~$=-2$ have an averaged \fescLyC\ $\lesssim5\%$ which is smaller than \fescLya\ $\lesssim10\%$ estimated from our linear fit (panel 0-1 of Figure \ref{propt}). \id7 is the bluest one in our sample with \fescLya\ $\sim50\%$. This galaxy and the luminous LAEs with very blue UV continua (\buv\ $\lesssim3$) reported by \citet{jiang20a} are supposed to be strong LyC leakers to contribute ionizing photons. They are thus excellent targets to carry out \jwst\ IR spectroscopic followup for the ionization lines.

\begin{figure*}[t]
\centering
\includegraphics[angle=0, width=0.75\textwidth]{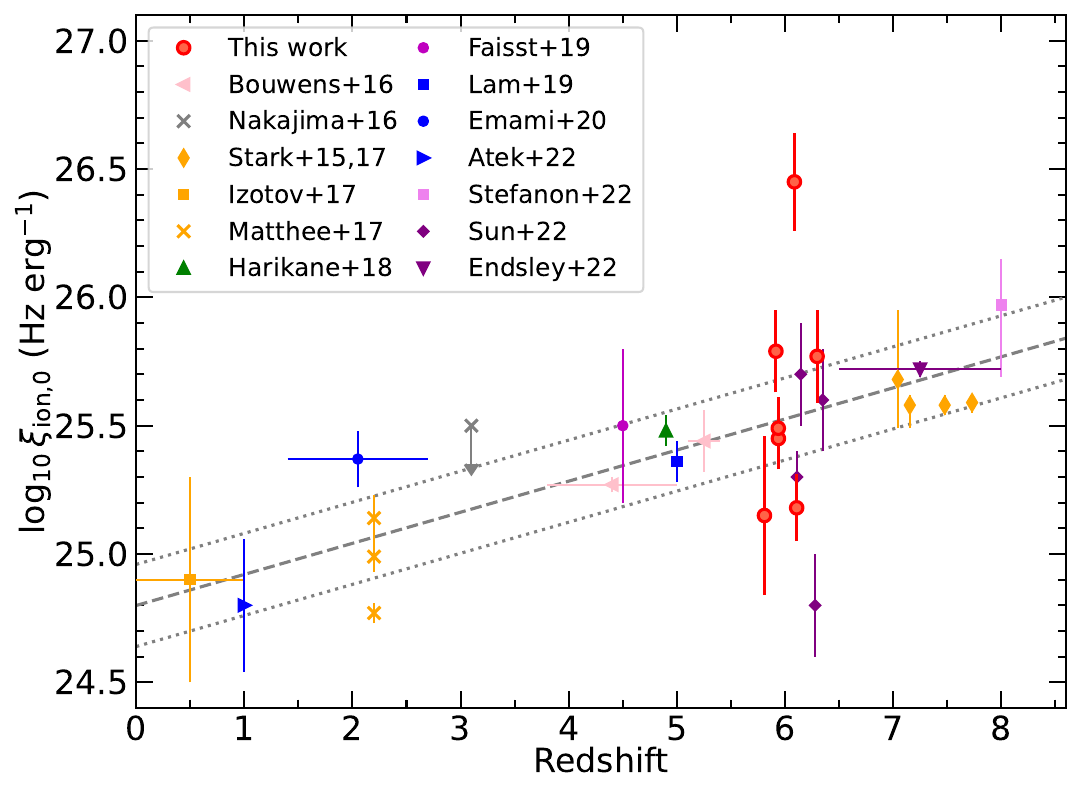}
\caption{Redshift evolution of \ksio\ for SF galaxies. The red circle symbols mark our results. Other symbols represent the literature studies including \citet{bouwens16}, \citet{nakajima16}, \citet{stark15}, \citet{stark17}, \citet{izotov17}, \citet{matthee17a}, \citet{harikane18b}, \citet{faisst19}, \citet{lam19}, \citet{emami20}, \citet{atek22}, \citet{stefanon22}, \citet{sun22b}, and \citet{endsley22}. The gray dashed line indicates the best-fit linear relation with a $1\sigma$ uncertainty region enclosed by the gray dotted lines.
\label{ksi}}
\end{figure*}

\subsection{\ksio-\lya\ Correlation}

The panel 3-2 (3-3) of Figure~\ref{propt} shows a potential correlation between the \ksio\ and \lya\ luminosity (\lya\ equivalent width) for the $z\simeq6$ sample in this work. We obtain a linear relation between \ksio\ and $L({\rm Ly\alpha})$:
\begin{equation}
 {\rm log_{10}}\,\xi_{\rm ion, 0}= (0.87 \pm 0.19)\,{\rm log_{10}}\,L({\rm Ly\alpha}) - (11.3 \pm 8.2),
 \label{eq3}
\end{equation} shown as a red dotted line in panel 3-2. \citet{2022arXiv221101351S} also found a similar trend in a sample of SF galaxies at $3\leq z\leq5$. We notice that such correlations does not always keep for the LAEs at $z\sim2$ from \citet{matthee21}, especially for those with relative bright \lya\ flux. A direct reason is that the $z\sim2$ LAEs in \citet{matthee21} generally have higher \fescLya, even reaching $\gtrsim1$. \citet{matthee17c} also obtained several galaxies have \fescLya\ $\gtrsim1$ and they explained it using the mechanisms including, for example, \lya\ emission is produced by a different mechanism (like cooling radiation; \citealt{dijkstra14}). We look forward to compare their results updated with the additional \ha\ information.

The \ksio-$L({\rm Ly\alpha})$ trend can be reasonably expected because harder ionizing radiation of more young stellar populations (higher sSFR shown by the panel 2-2) tend to produce more ionizing photons per non-ionizing photons \citep[e.g.,][]{balestra13, mainali17} and ionize the ISM more efficiently, eventually enhancing \lya\ production, transfer, and escape. Note that \ewlya\ can be inferred to be proportional to \ksi$\times$\fescLya\ \citep[e.g.,][]{harikane18b}. This trend implies that galaxies with \lgLya~$\gtrsim43.5$ (at the bright end of \lya\ luminosity function) may have a high ionizing photon production efficiency reaching log$_{10}$\,\ksio~$\gtrsim27$. Conversely speaking, SF galaxies with low \ksio\ (${\lesssim}25$) may tend to emit negligible \lya\ emission. If such a trend exists, the very (\lya) luminous galaxies could fully ionize their surrounding neutral \hi\ gas with a modest or even small \fescLyC\ value. This situation is also consistent in which the scenario that luminous \lya\ galaxies in the EoR power large ionizing bubbles \citep[e.g.,][]{zheng17, yajima18, hu21, ning22}.

\subsection{Implications of High \ksi}

We plot the redshift evolution of \ksio\ for the SF galaxies in Figure~\ref{ksi}. The red circle symbols indicate our results for the \lya\ galaxies at the end of EoR. The figure shows that the \ksio\ of SF galaxies increases with redshift. Among the seven \lya\ galaxies, three of them (\id1, 3, and 7) have high production efficiency of ionizing photons of log$_{10}$\,\ksio~$\gtrsim26$. This fact reveals the possibility that a portion of galaxies have high values of \ksi. \citet{maseda20} also found an elevated mean \ksi\ (log$_{10}$\,\ksi~$\gtrsim26$) in a sample of UV-faint and high-EW LAEs at $z\approx4-5$. \citet{finkelstein19} presented a reionization model dominated by low-mass and UV-faint galaxies, which would requires such high \ksi. Our results provide an observation evidence. The high values of \ksi\ have not been well explained in galaxy simulations. For example, \citet{wilkins16} explored the \ksi\ range of EoR galaxies in the BlueTides simulation, in which they used low-metallicity stellar population (with binary systems) models to obtain their highest \ksi\ of log$_{10}$\,\ksi~$\lesssim26$. \citet{yung20a} could not also provide such high \ksi\ with the semi-analytical models of galaxy formation. Note that their models only predict a rather weak dependence of \ksi\ with redshift, which means we have not well understood the physical processes responsible for the redshift evolution shown in Figure \ref{ksi} (the gray lines). In the meanwhile, a larger sample of fainter galaxies is necessary to be built to overcome the sampling bias at the higher redshift ($z>5$).

The median \ksio\ of our \lya\ galaxy sample is consistent the overall trend (gray lines). Previous results indicate EoR SF galaxies with \ksio\ on the overall trend can provide enough ionizing photon budget \citep[e.g.,][]{harikane18b, stefanon22}. Our sample implies that \lya\ galaxies may also play a similar role and contribute to the total balance of ionizing photons at $z\simeq6$. Note that the most luminous \lya\ galaxy in the sample, \id1 (identified by our Magellan M2FS survey), has a very high \ksio\ (log$_{10}$\,\ksio~$\sim26.5$) relative to other SF galaxies even though its \lya\ luminosity of \lgLya\ $\simeq43$ is only a little larger than the characteristic luminosity of \lya\ luminosity function \citep[e.g.,][]{hu10, kashikawa11, zheng17, ning22}. Its nature need to be revealed based on further spectroscopic observations. Our results support that \lya\ galaxies, especially those with intrinsically high \ewlya, may significantly contribute the ionizing photons during the EoR. We look forward to comparing our findings with the ionizing photon production efficiency estimated for other \lya-luminous galaxies \citep[from, e.g.,][]{ning20, ning22} which will be covered by the \jwst\ observations.

\section{Summary}
In this work, we present a pilot study of a spectroscopically confirmed sample of (\lya\ emitting) galaxies at redshift $z\simeq6$ based on the \jwst/NIRCam imaging data. The sample includes seven targets: one identified by our Magellan/M2FS spectroscopic survey and six observed by the previous CANDELSz7 survey. All the seven sources we analyze are Lyman break galaxies showing large differences in their \lya\ luminosity, ranging from no observed \lya\ line up to strong \lya\ line of \lgLya\ $\simeq43.3$. These objects have been covered by two \jwst/NIRCam imaging surveys, PRIMER and UDF-MB, which employ a serious of SW and LW bands.

We have obtained their \lya- and \ha-related properties by combining the NIRCam broad- and/or medium-bands. Based on the results, we also revealed the potential correlations among the properties of \lya\ galaxies at $z\simeq6$. We summarize our findings as follows: 
\begin{itemize}
 \item{6/7 galaxies have \lya\ escape fractions of \fescLya~$\lesssim10\%$ regardless of their \ewlya\ and \ewha, which might be the status for most of star-forming galaxies at $z\simeq6$.} 
 \item{One \lya\ galaxy which is relatively faint in the rest-frame UV (\muv~$\simeq-19$) with an extremely blue UV slope (\buv~$\simeq-3$) has a \lya\ escape fraction reaching \fescLya~$\simeq50\%$.}
 \item{\lya\ galaxies with fainter rest-frame UV continua and/or bluer UV slopes tend to have higher escape fraction of \lya\ photons (Equations [\ref{eq1}] \& [\ref{eq2}]).}
 \item{Our sample spread over a broad range of the ionizing photon production efficiency over log$_{10}$\,\ksio~$\sim25.0-26.5$ with a median value close to those of the galaxy samples at similar redshifts from other studies.}
 \item{Galaxies with more luminous \lya\ emission probably have higher production efficiency of ionizing photons (Equation [\ref{eq3}]).}
 \item{Our identified source (\id1 analyzed in this study) which is very luminous in \lya\ has a very high ionizing photon production efficiency of log$_{10}$\,\ksio\,(Hz erg$^{-1}$) $>26$. Its nature merits further investigation.}
\end{itemize}
Our results agree with the scenario in which \lya\ galaxies may serve as a significant contributor to cosmic reionization. The bluer and/or more luminous \lya\ galaxies are ideal targets for \jwst\ spectroscopic followup observations. We need a larger sample of \lya\ galaxies observed by \jwst\ for further analysis on the reionization sources.

\acknowledgments
We acknowledge support from the National Key R\&D Program of China (grant no.~2018YFA0404503), the National Science Foundation of China (grant no.~12073014, no.~11721303, and no.~11890693), and the science research grants from the China Manned Space Project with No.~CMS-CSST-2021-A05 and No.~CMS-CSST-2021-A07. We also thank the anonymous referee for the constructive comments and suggestions that improved this paper.

This work is based on the observations made with the NASA/ESA Hubble Space Telescope and NASA/ESA/CSA James Webb Space Telescope. The \hst\ observations are associated with the Cosmic Assembly Near-IR Deep Extragalactic Legacy Survey (CANDELS) program. \jwst\ data are obtained from the Mikulski Archive for Space Telescopes (MAST) at the Space Telescope Science Institute, which is operated by the Association of Universities for Research in Astronomy, Inc., under NASA contract NAS 5-26555 for \hst\ and NAS 5-03127 for \jwst. The \jwst\ observations are associated with programs GO-1837 and GO-1963.

\facilities{\hst\ (WFC3/IR), \jwst\ (NIRCam)}
\bibliography{ms.bbl}
\end{document}